\documentclass[aps,prl,twocolumn,english]{revtex4-2}
\usepackage{blindtext}

\usepackage{amsmath, amssymb,amsthm,dsfont,graphicx,mathtools,multirow,rotating,placeins,verbatim,dcolumn}
\usepackage[pdfborder={0 0 0}]{hyperref}

\usepackage{color}

\newcommand{\black}{\color{black}}

\newcommand{\A}{\mathcal{A}}
\newcommand{\F}{\mathcal{F}}
\newcommand{\tr}{\text{tr}}
\def\zb{\bar{z}}

\def\be{\begin{equation}}
\def\ee{\end{equation}}
\def\ba{\begin{eqnarray}}
\def\ea{\end{eqnarray}}
\def\bi{\begin{itemize}}
\def\ei{\end{itemize}}
\newcommand{\beq}{\begin{eqnarray}}
\newcommand{\eeq}{\end{eqnarray}}

\begin{document}
\title{A General Hierarchy of Charges at Null Infinity via the Todd Polynomials}
\author{Silvia Nagy$^a$}
\email{silvia.nagy@durham.ac.uk}
\author{Javier Peraza$^{b,c}$}
\email{javier.perazamartiarena@concordia.ca}
\author{Giorgio Pizzolo$^a$}
\email{giorgio.pizzolo@durham.ac.uk}
\affiliation{$^a$Department of Mathematical Sciences, Durham University, Durham, DH1 3LE, UK}
\affiliation{$^b$Facultad  de  Ciencias, Universidad  de  la  Rep\'ublica, 
Ig\'ua  4225,  Montevideo,  Uruguay}
\affiliation{$^c$Concordia University, Mathematics and Statistics Department, Montreal, Canada}

\begin{abstract}
We give a general procedure for constructing an extended phase space for Yang-Mills theory at null infinity, capable of handling the asymptotic symmetries and construction of charges responsible for sub$^n$-leading soft theorems at all orders. The procedure is coordinate and gauge-choice independent, and can be fed into the calculation of both tree and loop-level soft limits. We find a hierarchy in the extended phase space controlled by the Bernoulli numbers arising in Todd genus computations. We give an example of a calculation at tree level, in radial gauge, where we also uncover recursion relations at all orders for the equations of motion and charges.
\end{abstract}

\maketitle

\section{Introduction}

The null boundary of asymptotically flat spacetimes ($\mathcal{I}$) is an arena where unexpected deep connections between a priori different physical results can be made manifest, driving the quest for a flat space version \footnote{See \cite{Pasterski:2021raf,Raclariu:2021zjz} and references within.} of the celebrated holographic principle \cite{Maldacena:1997re}. A prime example of this is the duality between the soft theorems and the so-called large gauge transformations, via Ward identities \cite{Strominger:2013lka,Strominger:2013jfa,He:2014cra,Campiglia2015,Strominger2014a,Lysov:2014csa,Ferrari:1971at}. The soft theorems describe the behaviour of scattering amplitudes in the limit where the energy of one or more of the particles vanishes. They are by now an established result in QFT \cite{Weinberg1965,Low1958}. In contrast, the large gauge transformations are local symmetries, which, unlike their standard gauge counterparts, do not vanish at the boundary and, hence, have a dynamical role in the physical phase space. 

An essential requirement for formulating the above connection is the existence of a well-defined phase space at null infinity $\mathcal{I}$, on which the symmetries act canonically. It consists of the free data of the theory, and it allows us to construct the charges corresponding to the large gauge transformations, needed for the Ward identities to hold. The soft theorems can be formulated more precisely as an expansion in the small energy parameter, and the ideas above have been established at leading order in a variety of theories, as well as at some sub-leading orders \cite{Campiglia2020,Campiglia:2021oqz} \footnote{For an alternative approach, the so-called corner symmetry construction, see e.g. \cite{Donnelly:2016auv,Speranza:2017gxd,Freidel:2020ayo,Freidel:2020svx,Freidel:2020xyx,Freidel:2021dxw,Ciambelli:2021nmv,Freidel:2021cjp,Freidel:2021dfs,Ciambelli:2021vnn}}. An important question is whether we can extend this to all orders in the energy expansion, to give a full construction of the extended phase space. There are some encouraging signs coming from simplified set-ups, namely massless QED \cite{Peraza:2023ivy} and self-dual theories in light-cone gauge \cite{Nagy:2022xxs}.

In this paper we provide an affirmative answer to the question above, working in the context of Yang-Mills theory. We present a generalization of the Stueckelberg procedure \cite{Stueckelberg:1938hvi}, an algorithm normally used for restoring a broken gauge symmetry by the inclusion of a Goldstone-type field. This allows us to construct an extended phase space capable of accommodating the large gauge symmetries necessary for the charges corresponding to sub$^n$-leading soft theorems, for arbitrarily large $n$. Remarkably, the procedure works independently of coordinate and gauge choice, even accommodating field dependent gauge parameters. We also make no assumptions at this stage about the fall-off in the coordinate dual to the energy ($u$ in Bondi coordinates), which means it can be applied to loop-level soft theorems as well (see e.g. \cite{Sahoo:2018lxl, Pasterski:2022djr, Donnay:2022hkf, Agrawal:2023zea, Choi:2024ygx,Campiglia:2019wxe,AtulBhatkar:2019vcb}). 

Interestingly, we find that the symmetry transformations of the extended phase space Stueckelberg fields and hence the hierarchical relations for the sub$^n$-charges are controlled by the Bernoulli numbers, which arise from the perturbative expansion of (an operator version of) the characteristic power series generating the so-called Todd polynomial \cite{ToddPol}\footnote{We thank Hyungrok Kim for pointing out that the operator we define in \eqref{O_inv_def} is related to the Todd polynomial}.

In the second half of the paper, we show that the equations of motion satisfy the necessary recursion relations to all orders in the radial expansion, and how this relates to the energy expansion for sub$^n$-leading theorems. To our knowledge, this has not been presented in the literature at arbitrary order before. We accomplish this for various gauge choices and coordinates, as will be detailed in a longer companion paper \cite{Nagy:2024jua}. In this Letter, we will specialise to radial gauge in Bondi coordinates. 

Using the extended phase space constructed earlier, we derive the expressions for the sub$^n$-leading charges, living at null infinity. A standard renormalization procedure is carried out in order to avoid infrared divergences (see e.g. \cite{Peraza:2023ivy}). Restricting to tree-level, we also give a recursion relation for the charges, using the very same recursive relation from the equations of motion. For any $n\geq 0$, we prove the existence of a closed sub-algebra of charges, recovering known results from the literature at level $n=0$ (\cite{Strominger:2013lka}) and $n=1$ (\cite{Campiglia:2021oqz}).

In the context of Ward identities, certain higher derivative interactions can display so-called \textit{quasi-universal} contributions in addition to the universal terms \cite{Elvang:2016qvq, Laddha:2017vfh}. We compute these contributions to the sub-leading charge in Yang-Mills using the dressing procedure in our framework, for the particular example of an interaction of the form $\phi \tr ( F^2 )$ \footnote{See Supplemental Material at [] for the particular example of an interaction of the form $\phi \tr ( F^2 )$.}. 

The paper is structured as follows: first, we describe the general Stueckelberg procedure for constructing the extended phase space. We also give expressions for the charges and show that their algebra closes in the required way. Next, we give an explicit example by focusing on Bondi coordinates in radial gauge. We also give recursion relations for the equations of motion and charges at all perturbative orders in the radial and phase space expansions, establishing a consistent charge algebra at each order. Finally, we discuss the conclusions.  

We give a more detailed account of these calculations in the companion paper \cite{Nagy:2024jua}.

\section{Extended phase space to all orders in arbitrary gauge and coordinates}
\label{Extended phase space to all orders in arbitrary gauge and coordinates}
We are working in YM theory with the standard equation of motion in the absence of matter sources  
\be 
\mathcal{E}_\nu\equiv
\label{eom_gen}
D^\mu \mathcal{F}_{\mu\nu}=0,
\ee 
with 
\be
\mathcal{F}_{\mu\nu}=\partial_\mu \mathcal{A}_\nu-\partial_\nu \mathcal{A}_\mu - i [\mathcal{A}_\mu,\mathcal{A}_\nu] \ ,
\ee
and \black the gauge field transforming as:
\be\label{gauge_gen}
\mathcal{A}'_\mu=e^{i\Lambda}\mathcal{A}_\mu e^{-i\Lambda}+ie^{i\Lambda}\partial_\mu e^{-i\Lambda}.
\ee
We will assume that $\mathcal{A}_\mu$ satisfies some gauge condition
\be \label{gf_gen}
\mathcal{G}(\A_\mu)=0.
\ee
Let us denote our coordinates as $x^\mu=(\mathfrak{r},\vec{\mathbf{y}})$, where $\mathfrak{r}$ is our expansion parameter \footnote{Generally we will take $\mathfrak{r}$ to be the radial direction, but we will allow for other possibilities as well (see e.g. \cite{Nagy:2022xxs} for an example in light-cone gauge, where $\mathfrak{r}=V$).}. Let us allow for a very general expansion of the gauge field in terms of polyhomogeneous functions 
\be \label{A_st_exp}
\mathcal{A}_\mu=\sum_{n,k\geq0}A_\mu^{(-n;k)}(\vec{\mathbf{y}})\frac{\text{log}^k \mathfrak{r}}{\mathfrak{r}^n} \ ,
\ee 
such that $\lim_{\mathfrak{r}\to\infty}\frac{\text{log}^k \mathfrak{r}}{\mathfrak{r}^n}$ is of at most $\mathcal{O}(1)$. The logarithmic terms are necessary for certain gauge choices (see, e.g., \cite{Campiglia:2021oqz} in Lorenz gauge). Let us assume we have determined the radiative phase space:
\be 
\label{gen_ph_sp} 
\Gamma^0=\left\{\mathfrak{A}^{0}\ \text{satisfying \eqref{eom_gen} and \eqref{gf_gen}} \right\}. 
\ee
We now wish to allow for large gauge transformations with divergent behaviour as $\mathfrak{r} \rightarrow +\infty$, as expected in sub$^n$-leading soft limits:
\be \label{Lambda_plus_exp}
\Lambda_+(x)=\sum_{n,k}\mathfrak{r}^n \text{log}^k\mathfrak{r}\Lambda^{(n;k)}(\vec{\mathbf{y}}) \ ,
\ee
with $n$ and $k$ chosen such that $\mathfrak{r}^n \text{log}^k\mathfrak{r}$ diverges as $\mathfrak{r}\to\infty$. 

More generally, we could find a possibly field dependent parameter \footnote{This could arise as a result of imposing further constraints, see e.g. \cite{Campiglia:2021oqz} , where the field-dependent terms are a consequence of imposing Lorenz gauge.} given by
\be \label{gauge_par_field_dep}
\begin{aligned}
\breve{\Lambda}_+&= \breve{\Lambda}_+(\mathcal{A}_\mu(x),\Lambda_+(x))\\
&=\sum_{n,k}
\mathfrak{r}^n \text{log}^k\mathfrak{r}\ f^{(n;k)}(A_\mu^{(0;0)}(\vec{\mathbf{y}}),...,\Lambda^{(0;1)}(\vec{\mathbf{y}}),...) \ ,
\end{aligned}
\ee 
It is this composite object that is the starting point for the Stueckelberg procedure. We define the object
\be \label{comp_stueck}
\begin{aligned}
\breve{\Psi}&= \breve{\Psi}(\mathcal{A}_\mu(x),\Psi(x))\\
&=
\sum_{n,k}
\mathfrak{r}^n \text{log}^k\mathfrak{r}\ f^{(n;k)}(A^{(0;0)}_\mu(\vec{\mathbf{y}}),...,\Psi^{(0;1)}(\vec{\mathbf{y}}),...),
\end{aligned}
\ee
where $f$ is as in \eqref{gauge_par_field_dep}, in other words \eqref{comp_stueck} is obtained from \eqref{gauge_par_field_dep} via the replacement
\be \label{stueck_trick}
\Lambda_+(x)\quad \to \quad \Psi(x),
\ee 
where we have introduced our Stueckelberg field
\be \label{gen_psi_no_fields}
\Psi(x)=\sum_{n,k}\mathfrak{r}^n \text{log}^k\mathfrak{r}\Psi^{(n;k)}(\vec{\mathbf{y}}).
\ee 
with $n,k$ as in \eqref{Lambda_plus_exp}. We then claim that the extended phase space on which the sub-leading gauge transformations are well-defined is simply given by 
\be
\label{extended_phase_space_YM}
\Gamma_{\infty}^{\text{ext}} := \Gamma^0\times\{ \Psi(x) \text{ as defined in \eqref{gen_psi_no_fields}} \}. 
\ee
To verify this, we will derive the transformation rule for $\Psi(x)$ and use this to construct our sub$^n$-leading charges, which will act \textit{canonically} on $\Gamma_{\infty}^{\text{ext}}$. We first define the extended gauge field by applying \eqref{stueck_trick} to \eqref{gauge_gen}
\be\label{stuec_gen}
\tilde{\mathcal{A}}_\mu = e^{i\breve{\Psi}}\mathcal{A}_\mu e^{-i\breve{\Psi}}+ie^{i\breve{\Psi}}\partial_\mu e^{-i\breve{\Psi}}.
\ee 
We then require that $\tilde{\mathcal{A}}_\mu$ has a standard linearised gauge transformation
\be
\label{delta_Lambda_HatA}
\delta_{\breve\Lambda}\tilde{\mathcal{A}}_\mu = \tilde{D}_\mu\breve\Lambda  \ ,
\ee
where $\tilde{D}$ is defined w.r.t. $\tilde{\mathcal{A}}$ and
\be \label{lambda_split}
\breve\Lambda=\Lambda^{(0)}+\breve\Lambda_+ \ .
\ee 
where $\Lambda^{(0)}$ is the parameter for the leading order large gauge transformation. This requirement stems from the fact the Stueckelberg field should not be thought of as an additional field in the bulk, but rather as originating from the longitudinal components of the gauge field which were discarded in the expansion \eqref{A_st_exp}. Then from \eqref{delta_Lambda_HatA} and \eqref{stuec_gen}, and assuming that the transformation of the original gauge field $\A_\mu$ in \eqref{A_st_exp} is unchanged, we derive 
\be\label{psi_inv_O} 
\delta_{\breve\Lambda}\breve\Psi = \mathcal{O}_{-i\breve\Psi}^{-1}({\breve\Lambda}-e^{i\breve\Psi}\Lambda^{(0)}e^{-i\breve\Psi}),
\ee 
where we have introduced the operator
\be \label{O_def_gen}
\mathcal{O}_X  := \frac{1-e^{-ad_X}}{ad_X} = \sum_{k=0}^\infty\frac{(-1)^k}{(k+1)!}(ad_X)^k \ ,
\ee 
where $ad_X(Y)=[X,Y]$ and $\mathcal{O}_X^{-1}$ can be interpreted as an operator version of the characteristic power series generating the Todd polynomials \cite{ToddPol}
\be \label{O_inv_def}
\mathcal{O}_X^{-1}=\frac{ad_X}{1-e^{-ad_X}} = \sum_{m=0}^{\infty}\frac{B^+_m \left(ad_X\right)^m}{m!} \ ,
\ee
in terms of the Bernoulli numbers $B^+_m$. This allows us to extract the transformation of $\breve\Psi$ at order $m$ in $\breve\Psi$
\be \label{delta_m_Psi}
\delta_{\breve\Lambda}^{[m]}\breve\Psi = \frac{B_m^+}{m!}(ad_{-i\breve\Psi})^m\left[{\breve\Lambda}+(-1+2\delta_{m,1})\Lambda^{(0)}\right].
\ee
We remark that for $m>1$ and odd, the above vanishes, since odd Bernoulli numbers $B^+_{2k+1}$ vanish for $k>0$. Incidentally, we remark that, in view of the identity 
\be 
\zeta(m) = \frac{(-1)^{\tfrac{m}{2}+1}}{2} \frac{B_m^+ (2 \pi)^m}{m!}, \ \text{for} \ m \ \text{even},
\ee 
where $\zeta(m)$ is  the Riemann $\zeta$-function \cite{garfken67:math}, we can recast \eqref{delta_m_Psi}, for $m>1$, as
\be 
\delta_{\breve\Lambda}^{[m]}\breve\Psi = \frac{2 (-1)^{\tfrac{m}{2}+1}\zeta(m)}{(2 \pi)^m}(ad_{-i\breve\Psi})^m\left[{\breve\Lambda}+(-1+2\delta_{m,1})\Lambda^{(0)}\right].
\ee
Next, recall that $\breve\Lambda$ is a composite object depending on the Stueckelberg field $\Psi$ and possibly also the gauge field itself, with the explicit form of $f$ in \eqref{comp_stueck} determined by the gauge choice. Note that at $m=0$, $\breve{\Psi}$ will transform via a shift, as expected for a Goldstone-type mode \footnote{For other works where Goldstone modes appear in the context of soft limits and celestial holography, see e.g. \cite{Kampf:2023elx,Hamada:2017atr,Kapec:2022hih,Garcia-Sepulveda:2022lga,He:2024ddb}}.

The sub$^n$-leading charges, at all orders in $n$, arise as the natural (renormalized) generalization of the leading charge 
\be \label{gen_charge_gen}
\tilde{Q}_{\breve\Lambda} = \int_{\mathcal{B}^2} \tr ({\breve\Lambda} \tilde{\F}^{\mu \nu})^{ren} dS_{\mu \nu},
\ee
where we introduced the generalized field strength
\be \label{gen_field_str}
\tilde{\F}_{\mu \nu}=e^{i\breve\Psi}\F_{\mu\nu}e^{-i\breve\Psi}.
\ee
Such renormalization has already been worked out in the QED case at all orders \cite{Peraza:2023ivy}, and we will proceed in the same way for the YM case, with an explicit example given in \autoref{sec_charge_construction}. The volume element on the codimension two hypersurface $\mathcal{B}^2$ is
\be 
d S_{\mu\nu} = dx_B dy_B \sqrt{g_B}m_\mu n_\nu ,
\ee 
where $g_B$ is the induced metric on $\mathcal{B}^2$ and $m_\mu$, $n_\nu$ are unit vectors orthogonal to it. We will find it simpler to work with the charge densities,
\be 
\tilde q_{\breve\Lambda}=\tr\left(\sqrt{g_B} {\breve\Lambda} \tilde{\F}_{mn}\right)^{(ren)},
\ee 
where we use the notation 
\be 
\tilde{\F}_{mn}=\tilde{\F}^{\mu \nu}m_\mu n_\nu.
\ee
Finally, making use of \eqref{psi_inv_O}, and with the variation of the gauge field unchanged, we can explicitly compute the charge algebra. We note that it is deformed due to the presence of the field-dependent parameter, e.g. \cite{Barnich:2013sxa, Campiglia:2021oqz},
\be \label{commGenStart}
\{ \tilde q_{{\breve\Lambda}_1},\tilde q_{{\breve\Lambda}_2}\}_*=\frac{1}{2}\left[\delta_{{\breve\Lambda}_1}\tilde q_{{\breve\Lambda}_2}+\tilde q_{\delta_{{\breve\Lambda}_1}{\breve\Lambda}_2}-(1\leftrightarrow2)\right]
=\tilde q_{[{\breve\Lambda}_1,{\breve\Lambda}_2]_*},
\ee
where the deformed bracket is 
\be 
[{\breve\Lambda}_1,{\breve\Lambda}_2]_*=-i[{\breve\Lambda}_1,{\breve\Lambda}_2]+\delta_{{\breve\Lambda}_1}{\breve\Lambda}_2-\delta_{{\breve\Lambda}_2}{\breve\Lambda}_1.
\ee
Details of the derivation will be provided in \cite{Nagy:2024jua}.

\section{Recursive construction in radial gauge}
\label{Recursive construction in radial gauge}

In this section we will present a recursion relation which allows us to construct the components of the gauge field at any order. This will be crucial in the construction of the charges.

Let us specialise to Bondi coordinates, working in a neighbourhood of $\mathcal{I}^+$
\begin{equation}
    ds^2=-du^2-2dudr+2r^2\gamma dzd\bar{z},
\end{equation}
where $\gamma =\tfrac{2}{(1+z\zb)^2}$. For simplicity, let us assume that the gauge choice \eqref{gf_gen} allows for a consistent field expansion free of logarithmic terms in \eqref{A_st_exp}\footnote{This assumption is supported by the polynomial tree level fall-offs in the $u$ coordinate which are directly related to the fall-offs in $r$ in radial gauge, see section 3 of \cite{Nagy:2024jua}. Other gauges, such as harmonic gauge, cannot be consistent for a $1/r$-expansion, e.g. \cite{Campiglia:2021oqz}}, such that a general tensor ${\mathcal{T}^{\mu_1\dots}}_{\nu_1\dots}$ is expanded as 
\begin{equation} \label{eq:generic tensor expansion}
{\mathcal{T}^{\mu_1\dots}}_{\nu_1\dots}(u,r,z,\Bar{z})
 =\sum_{n\in\mathbb{N}}\frac{1}{r^n} {{T^{(-n)}}^{\mu_1\dots}}_{\nu_1\dots} (u,z,\Bar{z}).
\end{equation}

\subsection{General recursion relations to arbitrary order in $r$ in radial gauge}
\label{subsec:radial rec}
Let us now specialise to radial gauge, where $\mathcal{A}_r$ vanishes, together with the standard condition $A_u^{(0)}=0$. The equations of motion in radial gauge, at arbitrary order in the radial expansion, are given in \footnote{See Supplemental Material at []}. From these equations with $n=2$, we first get\footnote{Here we use the convention about symmetrization and antisymmetrization of indices where $T_{(\mu \nu)\alpha ...}$ and $T_{[\mu \nu] \alpha ...}$ are defined as $\frac{1}{2} \left( T_{\mu \nu \alpha ...} + T_{\nu \mu \alpha... } \right)$ and $\frac{1}{2} \left( T_{\mu \nu \alpha ...} - T_{\nu \mu \alpha ...} \right)$ respectively, for some tensor $T$. 
}
\be \label{rec_au_-1}
A_u^{(-1)}=2\gamma^{-1}\left(\partial_{(z}A_{\bar{z})}^{(0)}+i\partial_u^{-1}[\partial_uA_{(z}^{(0)},A_{\bar z)}^{(0)}]\right),
\ee
where 
\be 
\partial_u^{-1} f(u , z ,\zb) := \int_{-\infty}^{u}f(\mathfrak{u},z,\zb) d\mathfrak{u}.
\ee 
For $n\geq2$ we find:
\begin{align}
    \label{eq:A_u(-n) in radial gauge full Bondi}
    A_u^{(-n)}&=-\tfrac{2\gamma^{-1}}{n}\left(\partial_{(z}A_{\Bar{z})}^{(1-n)}+\sum_{k=1}^{n-1}\tfrac{ik}{1-n}[A^{(1+k-n)}_{(z},A_{\Bar{z})}^{(-k)}]\right).
\end{align}
We notice that, in general:
\begin{align}
    \label{A_u(-n) dependence in radial full Bondi}
    A_u^{(-n)}& \quad \mathrm{depends} \ \mathrm{on} \quad \{A_z^{(-k)},A_{\bar{z}}^{(-k)}\}_{k<n}.
\end{align}
Let us now focus on the remaining $A_z$ and $A_{\zb}$ components. After some algebra we obtain, for $n\geq1$: 
\begin{align}\label{rec_Az_-n}
    A_z^{(-n)}&=\frac{\partial_u^{-1}}{2}\bigg(\partial_{z}A_{u}^{(-n)}+(1-n)A_z^{(1-n)}+\tfrac{1}{n}\partial_z(\gamma^{-1}F^{(1-n)}_{z\bar{z}}) \nonumber \\
    &\quad -\tfrac{i}{n}\sum_{k=1}^{n}[A_z^{(k-n)},(2n-k)A_u^{(-k)}+\gamma^{-1}F^{(1-k)}_{z\Bar{z}}]\bigg).
\end{align}

By plugging \eqref{eq:A_u(-n) in radial gauge full Bondi} into the above, and in light of \eqref{A_u(-n) dependence in radial full Bondi}, we see that we have
\begin{align}
    \label{A_z(-n) dependence in radial full Bondi}
    A_z^{(-n)}& \quad \mathrm{depends} \ \mathrm{on} \quad \{A_z^{(-k)},A_{\bar{z}}^{(-k)}\}_{k<n},
\end{align}
with a completely analogous result for $A_{\zb}^{(-n)}$. Finally, from \eqref{A_u(-n) dependence in radial full Bondi} and \eqref{A_z(-n) dependence in radial full Bondi} we see that all the gauge components can be constructed recursively from $\{A_z^{(0)},A_{\bar{z}}^{(0)}\}$.

The above equations, together with all the expressions in the Supplemental Material, can be checked explicitly to arbitrarily high order using the Mathematica package developed by one of the authors, which will be presented in \footnote{G. Pizzolo, to appear}.

As we will see below, for the construction of charges we will be specifically interested in $F_{ur}^{(-n)}$, and we reproduce the recursion formula for this below for convenience, 
\be
    F_{ur}^{(-2)}=2\gamma^{-1}\left(\partial_{(z}A_{\bar{z})}^{(0)}+i\partial_u^{-1}[\partial_uA_{( z}^{(0)},A_{\bar z)}^{(0)}]\right),
\ee 
and
\be
    F_{ur}^{(-n)}=-2\gamma^{-1}\left(\partial_{(z}A_{\Bar{z})}^{(2-n)}+\sum_{k=1}^{n-2}\tfrac{ik}{2-n}[A^{(2+k-n)}_{(z},A_{\Bar{z})}^{(-k)}]\right), 
\ee
for $n\ge3$. Let us briefly discuss the consequences of the $\partial_u^{-1}$ operator appearing in the recursion relations above. We will assume that we are working at tree level, in which case at leading order we have \cite{Campiglia:2018dyi,Peraza:2023ivy}
\be 
\lim_{u\to\pm\infty}F_{ur}^{(-2)}(u,z,\zb)=F_{ur}^{(-2,0)}(z,\zb)+o(u^{-\infty}),
\ee 
where $F^{(-n,k)}_{ur}$ denotes the coefficient of $\frac{u^k}{r^n}$ in a formal expansion in $r$ and $u$, and $o(u^{-\infty})$ is a remainder that falls-off faster than $|u|^{-n}$, for any $n>0$. The recursion relations \eqref{rec_au_-1}, \eqref{eq:A_u(-n) in radial gauge full Bondi} and \eqref{rec_Az_-n} imply that the fields have the following polynomial expansion in $u$
\be 
\begin{aligned} 
\lim_{u\to\pm\infty}A_z^{(-n)}(u,z,\zb)&=\sum_{k=0}^{n}A_z^{(-n,k)}(z,\zb) u^k+o(u^{-\infty})\\
\lim_{u\to\pm\infty}A_u^{(-n)}(u,z,\zb)&=\sum_{k=0}^{n-1}A_u^{(-n,k)}(z,\zb) u^k+o(u^{-\infty})
\end{aligned}
\ee 

In the companion article \cite{Nagy:2024jua} we additionally present an arbitrary order recursion relation in light-cone gauge. 

\subsection{Charge construction and Algebra} \label{sec_charge_construction}
In radial gauge in Bondi coordinates, the renormalized charge \eqref{gen_charge_gen} reduces to 
\be 
\tilde{Q}_\Lambda = \int_{S^2} \tr \left( \sum_{l = 0}\Lambda^{(l)} \left( r^2 \tilde{F}_{r u} \right)^{(-l)} \right) dS_{S^2}, 
\ee
where $dS_{S^2} = \gamma dz d\zb$, and $\tilde{F}_{r u}$ is defined as in \eqref{gen_field_str}. We shall not impose any further constraints, thus leading to a simplified version of the expression for $\breve\Psi$, eqn. \eqref{comp_stueck}, and we have 
\be 
\breve\Psi(x)=\Psi(x)=\sum_{k=1}^{\infty}r^k\Psi^{(k)}(u,z,\zb).
\ee 
 For each $l\geq 0$, consider the \textit{extended} charge density associated to $\Lambda^{(l)}$, defined as 
\be 
\begin{aligned}
\tilde{q}&_{\Lambda^{(l)}}: = \tr \bigg( \Lambda^{(l)} F_{r u}^{(-2-l)} - [i\Psi^{(1)}, \Lambda^{(l)}] F_{r u}^{(-3-l)} + \\
& \Big(  \frac{1}{2} [i\Psi^{(1)}, [i\Psi^{(1)}, \Lambda^{(l)} ] ] - [i\Psi^{(2)}, \Lambda^{(l)}] \Big) F_{r u}^{(-4-l)}+  ... \bigg).
\end{aligned}
\ee
Observe that $\tilde{q}_{\Lambda^{(l)}}$ is linear in both $\Lambda^{(l)}$ and the coefficients $F_{r u}^{(-i-l)}$.

In what follows we will define a hierarchy of sub$^n$-charges, labeled by $j+l\geq 0$ in the set of sequences $\{ \{ \overset{j}{q}_{\Lambda^{(l)}} \}_{l\geq 0}\}_{j\geq0}$, where $\overset{n}{q}$ denotes that we are working up to order $n$ in $\Psi$ in the expansion of $q$. At each cut-off in the expansion of $\Psi$ we will construct a closed charge algebra, that approximates $\tilde{q}_{\Lambda^{(l)}}$. Each charge algebra will correspond to the radiative, linear, quadratic, ..., approximations.

Let us now expand in the field $\Psi$. At $0$th order we obtain the standard large gauge symmetry charges. Working up to linear order, we have
\beq \label{lin1}
\overset{1}{q}_{\Lambda^{(0)}} &=& \tr \left( \Lambda^{(0)} F_{r u}^{(-2)} -[i\Psi^{(1)}, \Lambda^{(0)}] F_{r u}^{(-3)}\right), \\
\label{lin2}
\overset{0}{q}_{\Lambda^{(1)}} &=& \tr \left( \Lambda^{(1)} F_{r u}^{(-3)} \right).
\eeq
We note that \eqref{lin1} and \eqref{lin2} agree with the expressions previously computed in \cite{Campiglia:2021oqz}. At order $n$ we have the expressions below, showcasing how charges at order $n$ in the Stueckelberg field can be constructed recursively from the lower order charges:
\beq 
\overset{0}{q}_{\Lambda^{(n)}}  &=& \tr \left( \Lambda^{(n)} F_{r u}^{(-2-n)} \right) , \\
\overset{1}{q}_{\Lambda^{(n-1)}}  &=&  \overset{0}{q}_{\Lambda^{(n-1)}} -  \overset{0}{q}_{[i\Psi, \Lambda^{(n-1)}]^{(n)}}, \\
&...&  \nonumber\\
\overset{n}{q}_{\Lambda^{(0)}}  &=& \overset{n-1}{q}_{\Lambda^{(0)}} - \sum_{k=1}^{n} \overset{0}{q}_{\left( \frac{1}{k!} ad^k_{i\Psi}(\Lambda^{(0)})\right)^{(n)}}.
\eeq
This hierarchy is schematically represented in Figure 1 in \footnote{See Supplemental Material at []}.

Finally, the charge algebra spanned by $\{\overset{0}{q}_{\Lambda^{(n)}} , ... ,\overset{n}{q}_{\Lambda^{(0)}} \}$ at the $n-$th level can be shown to close via the identities
\be 
\{\overset{k}{q}_{\Lambda_1^{(l)}} , \overset{j}{q}_{\Lambda_2^{(m)}} \} = \left\lbrace \begin{array}{lc}
\overset{k+j-n}{q}_{-i[\Lambda_1^{(l)}, \Lambda_2^{(m)}]} & \text{if } l+m \leq n \\
   0  &  \text{otherwise.}
\end{array}    
\right.
\ee
where $k+l = j+m = n$. As will be shown in the longer companion paper \cite{Nagy:2024jua}, a subset of these charges is equivalent to the recursion relations presented in \cite{Freidel:2023gue, Geiller:2024bgf}, where a sector of the charges was shown to obey the infinite-dimensional Yang-Mills analogue of the  $w_{1+\infty}$ algebra.

\section{Conclusions}
\label{Conclusions}

We have given the first construction of an enlarged phase space at null infinity to all orders, capable of producing the charges needed for understanding the symmetry origin of sub$^n$-leading effects, both at tree- and loop-level, in the context of YM theory. The first novel feature of our construction is that it employs a generalisation of the Stueckelberg procedure, which has so far appeared in seemingly unrelated studies of massive gauge theories and cosmology \cite{Stueckelberg:1938hvi,Henneaux:1989zc,Nagy:2019ywi,Bansal:2020krz}. The unifying principle is the presence of a local broken symmetry, though in our case this is subtly related to the radial expansion at null infinity.

We demonstrated the procedure by taking the radial gauge in Bondi coordinates as an example, further giving general $n$-th order recursion relations in this context, which facilitates the construction of charges to all orders, working at tree-level. A new insight here is that the recursion relations for the charges within the hierarchy are controlled by the Bernoulli numbers in the expansion of the generating power series for the Todd polynomials. Since $B_{2k+1}$ are vanishing for $k>0$, it seems that the higher levels are controlled by the even sub$^n$-leading charges. We will explore this further in future work.

A natural question is whether this extends to gravity, where work at the first few sub-leading orders already exists in some contexts (e.g., \cite{Cachazo:2014fwa, Campiglia:2021oqz, Freidel:2021dfs, Fuentealba:2022xsz}), see also results in the Newman-Penrose formalism \cite{Geiller:2024bgf}. An encouraging suggestion comes from the toy model calculation in \cite{Nagy:2022xxs}, where the self-dual sector of gravity was considered. Additionally, a straightforward relation was established there between YM and gravity, advancing the double copy program at the level of fields and symmetries. An exciting prospect opened up by our results here is whether this can be extended to the full YM and gravitational theories\footnote{For some previous work on double copy relations at null infinity, see also \cite{Campiglia:2021srh,Adamo:2021dfg,Ferrero:2024eva,Mao:2021kxq,Godazgar:2021iae}}. The construction presented in this Letter is particularly well-suited for a generalization to gravity, via the finite action of a ``Stueckelberg'' diffeomorphism $\xi$ acting on the metric $g$, 
\be 
\breve{g} = e^{\mathcal{L}_\xi} g.
\ee

A proper understanding of the role of classical large gauge symmetries in the quantum symmetries of the S-matrix is necessary for establishing a holographic principle in asymptotically flat spacetimes (e.g., \cite{Pasterski:2021raf,Raclariu:2021zjz} and references therein). The hierarchical structure of charge algebras encompasses, in a subsector, an infinite-dimensional algebra in the so-called \textit{corner} approach to gauge theories and gravity \cite{Freidel:2021ytz,Freidel:2023gue} and in the celestial holography program \cite{Guevara:2021abz,Strominger:2021mtt}. It would be interesting to see how they fit in our generalized framework. 

In relation to scattering amplitudes, the next step is to directly apply the ideas in this article to the calculation of sub$^n$-leading soft theorems, including the loop effects \cite{Bianchi:2014gla,Sahoo:2018lxl, Pasterski:2022djr, Donnay:2022hkf,Agrawal:2023zea,Choi:2024ygx,Campiglia:2019wxe,AtulBhatkar:2019vcb}, via the Ward identities \footnote{For recent results at the first sub-leading level see \cite{Choi:2024ygx}}. A simple set-up that looks promising as a starting point for going to arbitrary orders is the self-dual sector, which has the benefit of being one-loop exact, for both Yang-Mills and gravity \cite{Bern:1993qk,Mahlon:1993si,Bern:1996ja,Bern:1998xc,Costello:2021bah,Costello:2022wso,Monteiro:2022nqt}, and which preserves the infinite dimensional algebras above at loop-level \cite{Ball:2021tmb,Mago:2021wje,Banerjee:2023jne,Monteiro:2022xwq}.

\begin{acknowledgments}
We thank Miguel Campiglia for providing comments on an early version of the manuscript. We are also grateful to Marc Geiller, Hyungrok Kim, Alok Laddha and Sam Wikeley for useful discussions. We also want to thank the anonymous referees for their useful comments, which allowed us to connect our work with different parts of the vast literature in asymptotic symmetries and soft theorems. J.P. was partially funded by Fondo Clemente Estable Project FCE\_1\_2023\_1\_175902 and by CSIC Group 883174. G.P. is funded by STFC Doctoral Studentship 2023. S.N. is supported in part by STFC consolidated grant T000708.
\end{acknowledgments}

\end{document}